\documentclass[aps,prl,showpacs,superscriptaddress,twocolumn]{revtex4}

\usepackage{mathrsfs}
\usepackage{amsfonts}
\usepackage{amssymb}
\usepackage{amsmath}
\usepackage{graphicx}
\usepackage{color}

\newcommand{\ket}[1]{\vert #1 \rangle}
\newcommand{\bra}[1]{\langle #1 \vert}
\newcommand{\ketbra}[2]{\vert #1 \rangle \langle #2 \vert}

\newcommand{\norm}[1]{\| #1 \|}
\newcommand{\abs}[1]{| #1 |}

\newcommand{\tr}{\mathrm{Tr}}

\begin{document}

\title{Quantum Zeno Effect of General Quantum Operations}

\author{Ying Li}
\email{ying.li.phys@gmail.com}
\affiliation{Centre for Quantum Technologies, National University of Singapore, 3 Science Drive 2, Singapore 117543, Singapore}

\author{David Herrera-Marti}
\affiliation{Centre for Quantum Technologies, National University of Singapore, 3 Science Drive 2, Singapore 117543, Singapore}

\author{Leong Chuan Kwek}
\affiliation{Centre for Quantum Technologies, National University of Singapore, 3 Science Drive 2, Singapore 117543, Singapore}
\affiliation{Institute of Advanced Studies, Nanyang Technological University, 60 Nanyang View Singapore 639673, Singapore}
\affiliation{National Institute of Education, 1 Nanyang Walk Singapore 637616, Singapore}

\date{\today}

\begin{abstract}
In this paper, we show that the quantum Zeno effect occurs for any frequent quantum measurements or operations.
As a result of the Zeno effect, for non-selective measurements (or trace preserving completely positive maps), the evolution of a measurement invariant state is governed by an effective Hamiltonian defined by the measurements and the free-evolution Hamiltonian.
For selective measurements, the state may change randomly with time according to measurement outcomes, while some physical quantities (operators) still evolve as the effective dynamics.
\end{abstract}

\maketitle

\textit{Introduction}.---The phenomenon that frequent measurements can slow down the evolution of a quantum system is known as the quantum Zeno effect (QZE) \cite{Misra1977,Itano1990}.
As an interesting phenomenon in quantum physics, the QZE has been theoretically studied for decades and demonstrated in many experiments (see Ref. \cite{Facchi2008} for a review, and recent articles \cite{Smerzi2012,Wolters2013,OQZE1}).
If measurements project the state to the initial state, the state of a system can be totally frozen by the QZE.
Rather than freezing in the initial state, if measurements project the state to a multidimensional subspace that includes the initial state, the QZE allows the dynamics within the subspace, which is known as the quantum Zeno subspace \cite{Facchi2002}.
In the recent work Ref. \cite{OQZE1}, some of us proposed a new type of the QZE, which is called the \textit{operator} QZE.
In the operator QZE, the evolution of some physical quantities (operators) are frozen by frequent (non-commuting) measurements, while the quantum state may change randomly with time according to measurement outcomes.

In general, a quantum measurement corresponds to a set of measurement operators $\{ M_q \}$ satisfying the completeness equation $\sum _q M^{\dagger}_q M_q = \openone$ \cite{Nielsen}.
The post-measurement state for the measurement outcome $q$ is given by $\rho_q = p_q^{-1} M_q \rho M_q^\dag$, where $p_q = \tr (M_q \rho M_q^\dag)$ is the probability of the outcome $q$, and $\rho$ is the state of the system before the measurement.
If the measurement is non-selective, which means outcomes are not recorded, the measurement transforms the state as a trace preserving completely positive (CP) map $\mathcal{P} \rho = \sum _q M_q \rho M^{\dagger}_q$, where Kraus operators are measurement operators.
Actually, any trace preserving CP map can be formalised in the operator-sum representation \cite{Nielsen}.

In this paper, we show that the QZE occurs for any frequent quantum measurements or operations.
Under a frequently performed non-selective measurement (or trace preserving CP map), if the initial state is invariant under the measurement $\mathcal{P}$, the evolution is governed by an effective Hamiltonian defined by the measurement and the free-evolution Hamiltonian.
Under a frequently performed selective measurement, the state may change randomly with time according to measurement outcomes, but some operators still evolve as the effective dynamics.
In the effective dynamics, each \textit{measurement invariant subspace} (MIS), which is an irreducible common invariant subspace of measurement operators $\{M_q\}$, behaves like a single quantum state.
Actually, each set of isomorphic MISs contains a noiseless subsystem of the map $\mathcal{P}$ \cite{NS}, and the effective Hamiltonian drives the evolution of noiseless subsystems.
If there is not any non-trivial invariant subspace of $\{ M_q \}$ or MISs are not isomorphic with each other, the system is always totally frozen in the initial state.
The quantum Zeno subspace effect corresponds to the case that MISs are one-dimensional.

The most remarkable practical application of the QZE consists in suppressing decoherence and dissipation, which is crucial for practical quantum information processing.
The QZE can protect unknown quantum states in the Zeno subspace \cite{Vaidman1996}.
Recently, it is shown that the decoherence can be suppressed by the Zeno subspace effect while allowing for full quantum control \cite{Paz-Silva2012}.
Some of us proposed a protocol of protecting unknown quantum states form decoherence based on the operator QZE \cite{OQZE1}, which has the advantage over previous protocols that only two-qubit measurements rather than multi-qubit measurements are required.
In this paper, we find that in the manner of frequently performing a quantum measurement or operation that has isomorphic MISs, quantum information encoded in the noiseless subsystem associated with these isomorphic MISs can be protected while full quantum control is allowed.
Compared with generating noiseless subsystems with a sequence of pulses as in the theory of dynamical decoupling \cite{DD} and other QZE-based protocols, the QZE of general operations significantly enlarges the set of operations that can be engaged to protect quantum information.

\textit{Evolutions with non-selective measurements}.---We consider a system whose free evolution is governed by the Hamiltonian $H$.
The superoperator corresponding to the free time evolution is $\mathcal{U}(t) = e^{\mathcal{L}t} $, where the generator $\mathcal{L} \bullet = -i[H , \bullet]$.
As a typical model of the QZE, we suppose the measurement is performed $N$ times during the entire time of evolution $\tau$ at equal interval and each measurement is performed instantly, meaning that the measurement can be implemented in a negligible amount of time.
If the measurement is a non-selective measurement $\mathcal{P}$, the time evolution of the state reads \cite{Paz-Silva2012}
\begin{equation}
\rho(\tau) = [ \mathcal{P} \mathcal{U}(\tau/N) ]^N \rho(0).
\label{StaEvo}
\end{equation}
Here, the initial state is a \textit{measurement-invariant operator} (MIO), i.e., $\mathcal{P} \rho(0) = \rho(0)$.

For projective measurements \cite{Misra1977,Itano1990,Facchi2008} or weak projective measurements \cite{Peres1990,Paz-Silva2012}, a MIO state is a state in the Zeno subspace, and the dynamics is governed by an effective Hamiltonian $H_{\pi_Z} = \pi_ZH\pi_Z$ in the limit $N\rightarrow \infty$.
Here, $\pi_Z$ is the projector of the Zeno subspace.

As the main result of this paper, we will prove that, for any non-selective measurement $\mathcal{P}$, the state evolves driven by an effective Hamiltonian $\tilde{H}$ in the limit $N\rightarrow \infty$, i.e.,
\begin{equation}
\rho(\tau) = e^{\tilde{\mathcal{L}} \tau} \rho(0),
\end{equation}
where $\tilde{\mathcal{L}} \bullet = -i[\tilde{H} , \bullet]$.
Here, we assume that the Hilbert space of the system is finite-dimensional, and $\norm{H}_1=J$ and $\norm{\tilde{H}}_1=\tilde{J}$ are both finite, where $\norm{\bullet}_1$ denotes the trace norm of an operator.
We would like to remark that this result is also valid for any trace preserving CP maps.

In the following five sections, firstly we analyse MIOs with three sections, then the effective Hamiltonian is given, and after that we prove the QZE of non-selective measurements.

\textit{Measurement invariant subspaces}.---Before we discuss MIOs, we have to decompose the Hilbert space orthogonally as
\begin{equation}
\mathcal{H}= [ \bigoplus_j (\mathcal{H}^{(j)}_S \otimes \mathcal{H}^{(j)}_R) ] \oplus \mathcal{H}_C.
\label{decomposition}
\end{equation}
Here, $\mathcal{H}^{(j)}_S$ and $\mathcal{H}^{(j)}_R$ are spanned by $\{ \ket{\Phi^{(j)}_s} \}$ and $\{ \ket{\psi^{(j)}_r} \}$, respectively, and $\mathcal{H}_C$ is spanned by $\{ \ket{\phi^C_l} \}$.

Subspaces $\{ \mathcal{H}^{(j)}_S \otimes \mathcal{H}^{(j)}_R \}$ are invariant subspaces of $\{M_q\}$, and each of them is composed of a set of isomorphic MISs $\mathcal{H}^{(j)}_S \otimes \mathcal{H}^{(j)}_R = \bigoplus_s \mathcal{H}^{(j)}_s$.
Here, the MIS $\mathcal{H}^{(j)}_s$ is spanned by $\{ \ket{\Phi^{(j)}_s} \otimes \ket{\psi^{(j)}_r} | r=1,2,\cdots,d^{(j)}_R\}$, and $d^{(j)}_R$ is the dimension of the subsystem $\mathcal{H}^{(j)}_R$.
Each set of isomorphic MISs is maximized, i.e., $\mathcal{H}^{(j)}_s$ and $\mathcal{H}^{(j^\prime)}_{s^\prime}$ are isomorphic iff $j = j^\prime$.
Here, two MISs are isomorphic means $\{\pi^{(j)}_s M_q \pi^{(j)}_s\}$ and $\{\pi^{(j^\prime)}_{s^\prime} M_q \pi^{(j^\prime)}_{s^\prime}\}$ are the same up to a unitary transformation, where $\pi^{(j)}_s$ is the projector of the subspace $\mathcal{H}^{(j)}_s$.

The complement subspace $\mathcal{H}_C$ neither is nor has a non-trivial invariant subspace of $\{ M_q \}$, but is an invariant subspace of $\{ M_q^\dag \}$.
If the algebra generated by $\{ M_q \}$ is a $\dag$ algebra, $\mathcal{H}_C$ is always empty \cite{Davidson}.
In general, the algebra generated by $\{ M_q \}$ may not be a $\dag$ algebra, thus $\mathcal{H}_C$ could be non-empty (see Example 1).

If there is not any non-trivial invariant subspace of $\{ M_q \}$, the Hilbert space $\mathcal{H}$ is irreducible and the decomposition reads $\mathcal{H}= (\mathcal{H}^{(1)}_S \otimes \mathcal{H}^{(1)}_R) \oplus \mathcal{H}_C$, where $\mathcal{H}^{(1)}_S$ is one-dimensional and $\mathcal{H}_C$ is empty.

With the decomposition of the Hilbert space, measurement operators reads $M_q = \sum_j \pi^{(j)} M_q \pi^{(j)} + M_q \pi^C$, where $\pi^{(j)}$ ($\pi^C$) is the projector of the subspace $\mathcal{H}^{(j)}_S \otimes \mathcal{H}^{(j)}_R$ ($\mathcal{H}_C$).
Up to a unitary transformation, $\pi^{(j)} M_q \pi^{(j)} = \openone^{(j)}_S \otimes M^{(j)}_q$, where $\openone^{(j)}_S$ ($\openone^{(j)}_R$) is the identity operator of the subsystem $\mathcal{H}^{(j)}_S$ ($\mathcal{H}^{(j)}_R$), and $\{M^{(j)}_q\}$ are operators of the subsystem $\mathcal{H}^{(j)}_R$.
Due to the completeness equation of $\{ M_q \}$, $\{ M^{(j)}_q \}$ also obey the completeness equation $\sum_q M^{(j)\dag}_qM^{(j)}_q = \openone^{(j)}_R$.
Because each $\mathcal{H}^{(j)}_s$ is a MIS, $\mathcal{H}^{(j)}_R$ is irreducible, i.e., $\mathcal{H}^{(j)}_R$ does not have any non-trivial invariant subspace of $\{ M^{(j)}_q \}$.

Decomposing the Hilbert space in a form similar to Eq. (\ref{decomposition}) is generally used to study noiseless subsystems \cite{NS,DD}, in which the decomposition are usually based on the representation theory \cite{Davidson} of the $\dag$ algebra generated by $\{ M_q,M_q^\dag \}$ and the complement subspace is always empty.
In this paper, rather than consider the $\dag$ algebra generated by $\{ M_q,M_q^\dag \}$, we have to consider the algebra generated by $\{ M_q \}$ for the purpose of analysing MIOs.
Actually, for unital maps, the complement subspace is always empty, and previous results of noiseless subsystems based on the $\dag$ algebra generated by $\{ M_q,M_q^\dag \}$ \cite{NS,DD} can be applied here.
We would like to remark that each $S$ subsystem is a noiseless subsystem of the map $\mathcal{P}$ \cite{NS}.
However, not all noiseless subsystems are $S$ subsystems that correspond to isomorphic MISs.

\textit{The limit of the map $\mathcal{S}_{N}$}.---To ensure the existence of MIOs, we define a map $\mathcal{S}_{N} = (1/N)\sum _{m=1}^{N} \mathcal{P}^{m}$, which is a trace preserving CP map.
For any operator $A$ with a finite trace norm, $\mathcal{S}_{N} A$ converges to a MIO in the limit $N \rightarrow \infty$.
If the trace of $A$ is nonzero ($A$ is positive), $\mathcal{S}_{\infty} A = \lim_{N \rightarrow \infty}\mathcal{S}_{N} A$ is always a nonzero (positive) MIO.

One can prove the limit of the map $\mathcal{S}_{N}$ by noticing $\norm{\mathcal{S}_{N+1} A - \mathcal{S}_{N} A}_1 = (N+1)^{-1}\norm{\mathcal{P}^{N+1} A - \mathcal{S}_{N} A}_1 \leq 2(N+1)^{-1}\norm{A}_1$ and $\norm{\mathcal{P} \mathcal{S}_{N} A - \mathcal{S}_{N} A}_1 = N^{-1}\norm{\mathcal{P}^{N+1} A - A}_1 \leq 2N^{-1}\norm{A}_1$.
Here, $\mathcal{P}^{N+1}$ and $\mathcal{S}_{N}$ are both trace preserving CP maps, which do not increase the trace norm of a Hermitian operator.
The operator $A$ may not be a Hermitian operator but can be written as a linear superposition of two Hermitian operators $A+A^\dag$ and $-iA+iA^\dag$.

\textit{Measurement invariant operators}.---A MIO $A$ is a fixed point of the map $\mathcal{P}$.
If $\mathcal{P}$ is unital, i.e., $\mathcal{P}\openone = \openone$, $A$ commutes with $\{ M_q,M_q^\dag \}$ \cite{FP}.
In this paper, we show that for a general map $\mathcal{P}$, $A$ can always be written as
\begin{equation}
A = \bigoplus_j (A^{(j)}_S \otimes \Lambda^{(j)}_R),
\label{MIO}
\end{equation}
where $A^{(j)}_S$ is an operator of the subsystem $\mathcal{H}^{(j)}_S$, and $\Lambda^{(j)}_R = (1/d^{(j)}_R) \mathcal{S}^{(j)}_{\infty} \openone^{(j)}_R$ is a MIO of the subsystem $\mathcal{H}^{(j)}_R$.
Here, $\mathcal{S}^{(j)}_{\infty} = \lim_{N \rightarrow \infty} (1/N) \sum_{m=1}^{N} \mathcal{P}^{(j)m}$ and $\mathcal{P}^{(j)} \bullet = \sum _q M^{(j)}_q \bullet M^{(j)\dagger}_q$ are maps of the subsystem $\mathcal{H}^{(j)}_R$.
If $\mathcal{P}^{(j)}$ is unital, $\Lambda^{(j)}_R = (1/d^{(j)}_R)\openone^{(j)}_R$.

To prove Eq. (\ref{MIO}), firstly, we consider Hermitian MIOs.
In the Supplementary Material \cite{SupMat}, we prove that a Hermitian MIO $A$ satisfies $\pi^C A = \pi^{(j)} A \pi^{(j^\prime)} =0$ for $j \neq j^\prime$ \cite{SupMat}, i.e., $A = \sum_j \pi^{(j)} A \pi^{(j)}$.
Because each $\pi^{(j)} A \pi^{(j)}$ is an operator in the invariant subspace $\mathcal{H}^{(j)}_S \otimes \mathcal{H}^{(j)}_R$, each $\pi^{(j)} A \pi^{(j)}$ is a Hermitian MIO.
In Ref. \cite{SupMat}, we also prove that $\Lambda^{(j)}_R$ is the unique Hermitian MIO of the measurement $\mathcal{P}^{(j)}$ up to a scalar factor.
Therefore, $\pi^{(j)} A \pi^{(j)}$ is proportional to $\Lambda^{(j)}_R$, i.e., $\pi^{(j)} A \pi^{(j)} = A^{(j)}_S \otimes \Lambda^{(j)}_R$, and the Hermitian MIO $A$ can also be written in the form of Eq. (\ref{MIO}).

If $A$ is a MIO but not Hermitian, $A+A^\dag$ and $-iA+iA^\dag$ are two Hermitian MIOs that can be written in the form of Eq. (\ref{MIO}).
Therefore, any MIO can be written in the form of Eq. ({\ref{MIO}}).

Now, we would like to show how to decompose the Hilbert space as Eq. (\ref{decomposition}).
To decompose the Hilbert space, one can consider the MIO $\Lambda = \mathcal{S}_{\infty} \openone = \bigoplus_j (I^{(j)}_S \otimes \Lambda^{(j)}_R)$, where $I^{(j)}_S \geq d_R \openone_S$ is an invertible Hermitian operator of the subsystem $\mathcal{H}^{(j)}_S$.
Because $\Lambda^{(j)}_R$ is also invertible \cite{SupMat}, the complement subspace $\mathcal{H}_C$ is spanned by eigenstates of $\Lambda$ with zero eigenvalues.
Then, one can decompose the Hilbert space as Eq. (\ref{decomposition}) by applying the representation theory \cite{Davidson} of the $\dag$ algebra generated by $\{ \pi^{SR} M_q \pi^{SR},\pi^{SR} M_q^\dag \pi^{SR} \}$ to the subspace spanned by eigenstates of $\Lambda$ with nonzero eigenvalues.
Here, $\pi^{SR} = \openone - \pi^C$ is the projector of the subspace spanned by nonzero-valued eigenstates.
Actually, one can prove that, the subspace spanned by zero-valued eigenstates of $\Lambda$ neither is nor has a non-trivial invariant subspace of $\{ M_q \}$, and each irreducible invariant subspace of $\{ \pi^{SR} M_q \pi^{SR},\pi^{SR} M_q^\dag \pi^{SR} \}$ is an irreducible invariant subspace of $\{ M_q \}$.

\textit{The effective Hamiltonian}.---The effective Hamiltonian reads
\begin{equation}
\tilde{H} = \bigoplus_j (\tilde{H}^{(j)}_S \otimes \openone^{(j)}_R),
\label{dualMIO}
\end{equation}
where $\tilde{H}^{(j)}_S = \tr_R [ \pi^{(j)} H \pi^{(j)} (\openone^{(j)}_S \otimes \Lambda^{(j)}_R) ]$ is a Hermitian operator of the subsystem $\mathcal{H}^{(j)}_S$.
As shown in Ref. \cite{SupMat}, the effective Hamiltonian satisfies $\mathcal{S}_{\infty}H\rho = \tilde{H}\rho$ and $\mathcal{S}_{\infty}\rho H = \rho \tilde{H}$ for any MIO state $\rho$.
Operators that can be written in the form of Eq. (\ref{dualMIO}) is called a dual MIO.

Driven by the effective Hamiltonian, the state initialized in a MIO state $\rho(0) = \bigoplus_j (\rho^{(j)}_S(0) \otimes \Lambda^{(j)}_R)$ evolves as $\rho(t) = \bigoplus_j (\rho^{(j)}_S(t) \otimes \Lambda^{(j)}_R)$, where $\rho^{(j)}_S(t) = e^{-i\tilde{H}^{(j)}_St} \rho^{(j)}_S(0) e^{i\tilde{H}^{(j)}_St}$.
Here, $\rho(t)$ is always a MIO.

If the Hilbert space is irreducible, there is only one MIO $\Lambda$ up to a scalar factor.
In this case, the state is frozen in $\Lambda$ as a result of the QZE.
Similarly, if $S$ subsystems are all one-dimensional, i.e., MISs are not isomorphic with each other, the system is always frozen in the initial state.

For projective measurements, one can find that the effective Hamiltonian coincides with the one predicted by the Zeno subspace theory (see Example 2).
For unital maps, the effective Hamiltonian $\tilde{H}^{(j)}_S = (1/d^{(j)}_R) \tr_R (\pi^{(j)} H \pi^{(j)})$, which is the same as the one generated by a sequence of pulses as in the theory of dynamical decoupling \cite{DD} (see Example 3).

\textit{Zeno effect of non-selective measurements}.---To show the effective dynamics, we suppose even in a very short time $\tau/N_{2}$, a large amount of ($N_{1}$) measurements are performed, i.e., $N=N_{1}N_{2}$, where $N_{1}$ and $N_{2}$ are both large numbers.
Firstly, we consider the time evolution of the first time interval of $\tau/N_{2}$, $\rho(\tau/N_{2}) = [ \mathcal{P} \mathcal{U}(\tau/N) ]^{N_1} \rho(0)$.
After expanding the free evolution superoperator $\mathcal{U}(\tau/N)$, we have
\begin{equation}
\rho(\tau/N_{2}) \simeq [ \mathcal{P}^{N_1} + (\tau/N_2)\mathcal{T}_{N_1} ] \rho(0),
\end{equation}
where $\mathcal{T}_{N_1} = (1/N_1)\sum _{m=1}^{N_1} \mathcal{P}^m \mathcal{L} \mathcal{P}^{(N_1-m)}$.
Because the initial state $\rho(0)$ is a MIO,
\begin{equation}
\rho(\tau/N_{2}) \simeq [ 1 + (\tau/N_{2}) \mathcal{S}_{N_1} \mathcal{L} ] \rho(0).
\end{equation}
As we have shown, if $N_1$ is large enough, $\mathcal{S}_{N_1} \mathcal{L} \rho(0) \simeq \tilde{\mathcal{L}} \rho(0)$, and
\begin{equation}
\rho(\tau/N_{2}) \simeq [ 1 + (\tau/N_{2})\tilde{\mathcal{L}} ] \rho(0)
\simeq e^{\tilde{\mathcal{L}} \tau/N_2} \rho(0),
\end{equation}
where the right side is a MIO.
For subsequent time intervals, we have similar conclusions.
Therefore, $\rho(\tau) \simeq e^{\tilde{\mathcal{L}} \tau} \rho(0)$.

A rigorous analysis \cite{SupMat} shows that $\rho(\tau) = e^{\tilde{\mathcal{L}} \tau} \rho(0) + \Delta$, where
\begin{equation}
\norm{\Delta}_1 \leq (\delta_{H} + \delta_{\tilde{H}} + \delta).
\label{bound}
\end{equation}
Here, we have $\delta_{H} = N_2 [ e^{2J\tau/N_2} - (1 + 2J\tau/N_2) ]$, similarly $\delta_{\tilde{H}} = N_2 [ e^{2\tilde{J}\tau/N_2} - (1 + 2\tilde{J}\tau/N_2) ]$, and $\delta = (\tau/N_2) \sum_{n=1}^{N_2} \norm{(\mathcal{S}_{N_1} \mathcal{L} -\tilde{\mathcal{L}}) \rho_n}_1$, where $\rho_n = e^{\tilde{\mathcal{L}} (n-1)\tau/N_2} \rho(0)$ is a MIO.
Without loss of generality, we set $N_1,N_2 = \sqrt{N}$.
Then, in the limit $N\rightarrow \infty$, all of $\delta_{H}$, $\delta_{\tilde{H}}$, and $\delta$ vanish.

\textit{Example 1: Decay channel}.---We consider a system with three states $\ket{g1}$, $\ket{g2}$, and $\ket{e}$.
Measurement operators are $M_1 = \ketbra{g1}{g1}+\ketbra{g2}{g2}$, $M_2 = (1/\sqrt{2})\ketbra{g1}{e}$, and $M_3 = (1/\sqrt{2})\ketbra{g2}{e}$.
In this example, $\mathcal{H}^C$ is non-empty and only includes the state $\ket{e}$, and $\ket{g1}$ and $\ket{g2}$ form two isomorphic one-dimensional MISs, respectively.
Any state initialized in the subspace spanned by $\ket{g1}$ and $\ket{g2}$ is a MIO.
As a result of the QZE, the evolution of such an initial state is frozen in the subspace.

\textit{Example 2: Zeno subspace}.---For a projective measurement $\mathcal{P}\bullet = \sum_j \pi^{(j)} \bullet \pi^{(j)}$, each common eigenstate of $\{\pi^{(j)}\}$ forms a one-dimensional MIS, and states in the same subspace $\pi^{(j)}$ are isomorphic.
In this example, $\tilde{H} = \sum_j \pi^{(j)}H\pi^{(j)}$.
If the state is initialized in the subspace $\pi^{(j)}$, the evolution is driven by the effective Hamiltonian $\pi^{(j)}H\pi^{(j)}$, which coincides with the Zeno subspace theory \cite{Facchi2002}.

\textit{Example 3: Symmetrizing operation}.---A symmetrizing operation \cite{DD} reads $\mathcal{P}\bullet = (1/\abs{G}) \sum_{g\in G} g \bullet g^\dag$, where $G$ is a group and $\abs{G}$ is the number of group elements.
In the theory of dynamical decoupling, the symmetrizing operation describes the effect of a sequence of pulses used for generating noiseless subsystems.
Here, the symmetrizing operation is supposed to be implemented as a general measurement (or trace preserving CP map).
In this example, MISs could be multi-dimensional if the group has multi-dimensional irreducible representations (is non-Abelian), and the effective Hamiltonian $\tilde{H} = \mathcal{P} H$.

\textit{Zeno effect of selective measurements}.---If measurement outcomes are recorded, the finial state $\rho (\tau; \{ q \})$ depends on all measurement outcomes $\{ q \}$ during the entire evolution.
The final state may not be a MIO.
And even if the driven Hamiltonian $H$ is absent, the state may change according outcomes during the evolution.
In the limit $N \rightarrow \infty$, the evolution of the state with selective measurements reads $\rho (\tau; \{ q \}) = \bigoplus_j [\rho^{(j)}_S(\tau) \otimes \rho^{(j)}_R(\{ q \})]$, where $\rho^{(j)}_S(\tau)$ is the state of the $S$ subsystem that evolves driven by the effective Hamiltonian, and $\rho^{(j)}_R(\{ q \})$ is the state of the subsystem $\mathcal{H}^{(j)}_R$ depending on measurement outcomes \cite{SupMat}.
We would like to remark that, because $\tr \rho^{(j)}_R(\{ q \})$ depends on measurement outcomes, the probability of the state in the subspace $\mathcal{H}^{(j)}_S \otimes \mathcal{H}^{(j)}_R$, $\tr [\rho^{(j)}_S(\tau) \otimes \rho^{(j)}_R(\{ q \})]$, depends on measurement outcomes.

\textit{Operator quantum Zeno dynamics}.---If the initial state is a product state of two subsystems, $\rho (0) = \rho^{(j)}_S(0) \otimes \Lambda^{(j)}_R$, the state is always confined in the subspace $\mathcal{H}^{(j)}_S \otimes \mathcal{H}^{(j)}_R$, i.e. $\rho (\tau) = \rho^{(j)}_S(\tau) \otimes \Lambda^{(j)}_R$ for the non-selective-measurement QZE and $\rho (\tau; \{ q \}) = \rho^{(j)}_S(\tau) \otimes \rho^{(j)}_R(\{ q \})$ for the selective-measurement QZE.
In this case, for a dual MIO $B = \bigoplus_j (B^{(j)}_S \otimes \openone^{(j)}_R)$, we have $\tr [\rho(\tau) B] = \tr [\rho (\tau; \{ q \}) B]$.
Therefore, for product-state initial states, we can define the effective evolution of operators $B(\tau) = e^{-\tilde{\mathcal{L}}\tau} B$, so that for both selective and non-selective measurements $\tr [\rho(\tau) B] = \tr [\rho (\tau; \{ q \}) B] = \tr [\rho(0) B(\tau)]$.

\begin{figure}[tbp]
\includegraphics[width=8.0 cm]{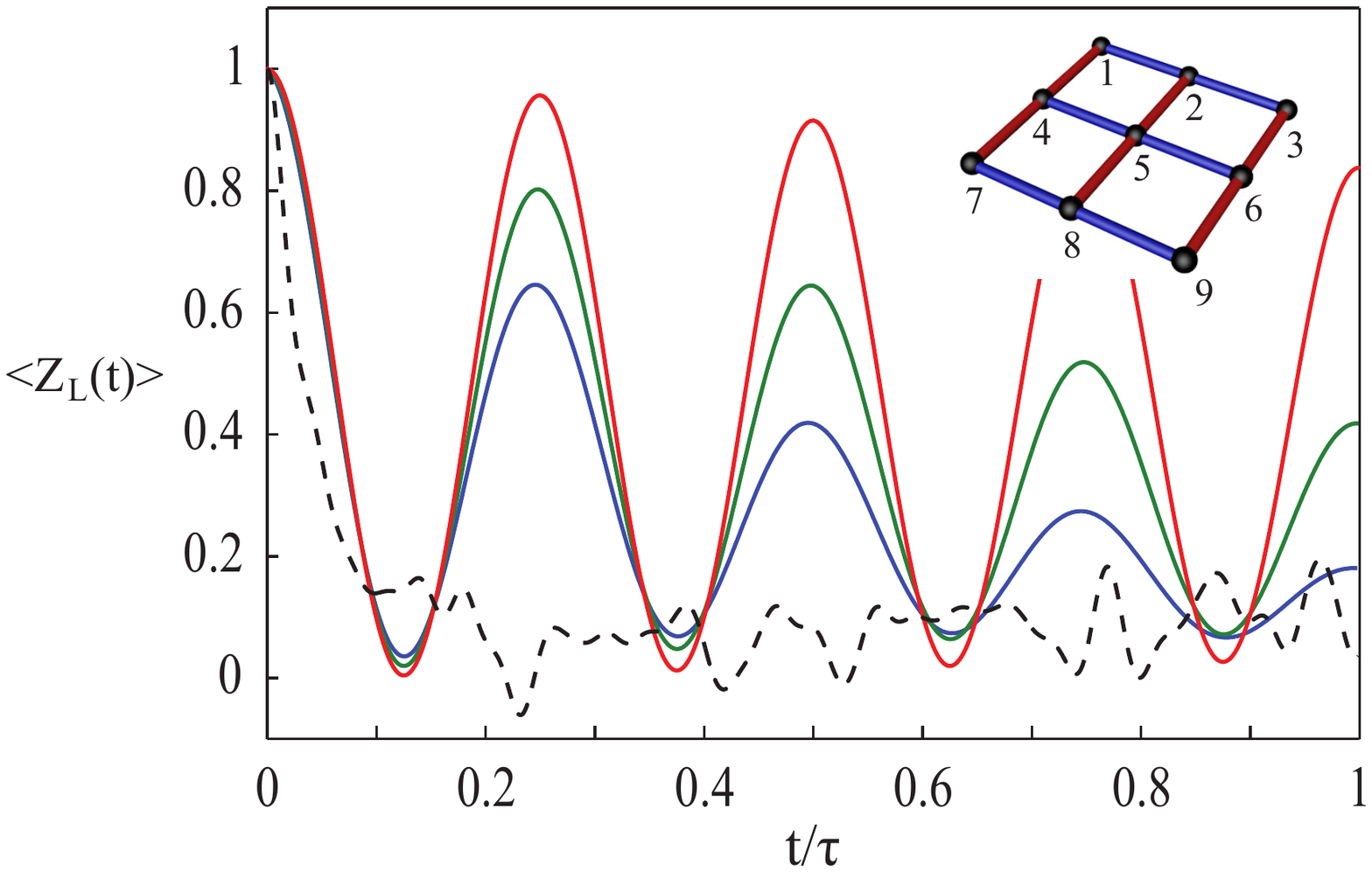}
\caption{
The expected value of the logical operator $Z_L$ of the $3 \times 3$ Bacon-Shor code.
In the inserted figure, each black round represents a physical qubit, and each (blue or red) bond represents a gauge operator.
The overall measurement is constructed by projectively measuring blue gauge operators first and then red gauge operators, and is performed $N$ times during the entire time of evolution $\tau$ at equal interval.
Here, the dashed black, solid blue, green, and red lines correspond $N = 0, 500, 1000$, and $5000$, respectively.
In this simulation, totally $8$ Hadamard gates are performed.
}
\label{Hadamard}
\end{figure}

\textit{Zeno quantum memory with general measurements}.---An important application of the QZE is protecting quantum states from decoherence \cite{Vaidman1996,Paz-Silva2012,OQZE1}.
In general, the free evolution of a quantum memory is governed by a Hamiltonian $H = H_0 + H_{\text{noise}}$, where the control Hamiltonian $H_0$ drives the evolution of the stored quantum state, and the noise Hamiltonian $H_{\text{noise}}$ induces decoherence due to the coupling with the environment.
If a measurement $\mathcal{P}$ has a multi-dimensional $S$ subsystem, e.g., $\mathcal{H}^{(1)}_S$, the quantum state stored in the subsystem $\mathcal{H}^{(1)}_S$ can be protected from decoherence by frequently performing the measurement $\mathcal{P}$ when the corresponding effective noise Hamiltonian $\tilde{H}^{(1)}_{S,\text{noise}} \propto \openone^{(1)}_S$.
Here, $\tilde{H}^{(1)}_{S,\text{noise}} = \tr_R [ \pi^{(1)} H_{\text{noise}} \pi^{(1)} (\openone^{(1)}_S \otimes \Lambda^{(1)}_R) ]$.
With a control Hamiltonian satisfying $\pi^{(1)}H_0\pi^{(1)} = \tilde{H}^{(1)}_{S,0} \otimes \openone^{(1)}_R$, the evolution of the stored quantum state is governed by $\tilde{H}^{(1)}_{S,0}$.
Therefore, the stored quantum state can be fully controlled.

\textit{Example 4: Bacon-Shor code}.---To illustrate the quantum control and the protection on a logical qubit encoded in a $S$ subsystem, we consider the $3 \times 3$ Bacon-Shor code \cite{Poulin2005,Bacon2006} (see the inserted figure of Fig. \ref{Hadamard}) as an example.
For the $3 \times 3$ Bacon-Shor code, only one logical qubit is encoded in $9$ physical qubits and the Hilbert space can be decomposed as $\mathcal{H} = \mathcal{H}_L \otimes \mathcal{H}_G$, where $\mathcal{H}_L$ is the Hilbert space of the logical qubit, and $\mathcal{H}_G$ is the Hilbert space of $8$ gauge qubits.
Logical Pauli operators are $Z_L = \sigma^z_2 \sigma^z_5 \sigma^z_8$ and $X_L = \sigma^x_4 \sigma^x_5 \sigma^x_6$, where $\sigma^z_i$ and $\sigma^x_i$ are Pauli operators of the $i$th physical qubit.
In the inserted figure, each blue (red) bond represents a gauge operator $\sigma^z_i \sigma^z_j$ ($\sigma^x_i \sigma^x_j$).

The idea of using the QZE to protected logical qubits of the Bacon-Shor code is firstly mentioned in Ref. \cite{Paz-Silva2012}.
By frequently measuring gauge operators, decoherence induced by one-local and two-local noises can be suppressed \cite{OQZE1}.
Hence, we employ the measurement $\mathcal{P} = \cdots \mathcal{P}_{c_2} \mathcal{P}_{c_1}$ to protect the logical qubit, where $c_1, c_2, \ldots$ are gauge operators.
The measurement of the gauge operator $c$ reads $\mathcal{P}_{c}(\zeta)\bullet =[(1+\zeta)/2]\bullet + [(1-\zeta)/2]c \bullet c$, where $0 \leq \zeta < 1$.
These two-qubit measurements can be implemented with two-qubit noisy interactions \cite{OQZE1}.
When $\zeta =0$ the measurement $\mathcal{P}_{c}(\zeta)$ is a projective measurement, and when $\zeta >0$ the measurement $\mathcal{P}_{c}(\zeta)$ corresponds to a weak measurement \cite{Aharonov1990,Brun2002,Oreshkov2005}.
Weak measurements can protect quantum states, which has been proved in protocols based on the Zeno subspace \cite{Paz-Silva2012}, while the evidence have been found numerically for the protocol based on the operator QZE \cite{OQZE1}.

For the measurement $\mathcal{P}$, the subsystem $\mathcal{H}_L$ and the subsystem $\mathcal{H}_G$ correspond to a $S$ subsystem and a $R$ subsystem, respectively.
Because $\mathcal{P}$ is unital, any MIO can be written as $A = A_L \otimes \openone_G/32$, and the effective Hamiltonian reads $\tilde{H} = (\tr_G H) \otimes \openone_G/32$.
For logical operators, $(\tr_G Z_L) \otimes \openone_G/32 = Z_L$ and $(\tr_G X_L) \otimes \openone_G/32 = X_L$.
For any one-local and two-local Pauli operators, $\tr_G \sigma^{\alpha}_i = \tr_G (\sigma^{\alpha}_i \sigma^{\beta}_j) =0$.

As an example, we consider performing Hadamard gates via the control Hamiltonian $H_0 = (\omega/\sqrt{2})(Z_L + X_L)$,
and the decoherence is induced by the noise Hamiltonian
\begin{equation}
H_{\text{noise}} = \omega ( \sum_i \sum_{\alpha = x,y,z} \sigma^{\alpha}_i
+ \sum_{(i,j)} \sum_{\alpha,\beta = x,y,z} \sigma^{\alpha}_i \sigma^{\beta}_j ),
\end{equation}
where the first (second) term corresponds to one-local (two-local) noises, and $(i,j)$ are two neighbouring qubits.
By frequently measuring gauge operators, decoherence of the logical qubit can be suppressed while logical operations (Hadamard gates) are performed, as shown in Fig. \ref{Hadamard}.

\textit{Discussions}.---In this paper, we have shown that the QZE occurs for any frequent quantum measurements or operations.
The time scale for implementing measurements has to be considered in future works, while in this paper measurements are supposed to be performed instantly.
We used the trace norm rather than the operator norm to describe the Hamiltonian strength.
Although for the finite-dimensional Hilbert space, a finite trace norm implies a finite operator norm for Hermitian operators, using the operator norm may be helpful in improving the bound in Eq. (\ref{bound}).
Besides suppressing decoherence, there are many other potential applications of the QZE \cite{Bernu2008,Alvarez2010,Bretschneider2012,Wen2012,McCusker2013, Erez2008,Maniscalco2008,Xu2009,Raimond2010,Huang2012}.

\begin{acknowledgments}
Y.L., D.H.M., and L.C.K. acknowledge support from the National Research Foundation \& Ministry of Education, Singapore.
We thank Paolo Zanardi and Sai Vinjanampathy for helpful discussions.
\end{acknowledgments}

\newpage
\widetext

\begin{center}
{\huge Appendix}
\end{center}

\section{Measurement invariant operators}

Firstly, we prove a lemma that is very useful for our discussions about MIOs.
We consider a Hermitian MIO $A$ in a Hilbert space that can be decomposed as $\mathcal{H} = \mathcal{H}^X \oplus \mathcal{H}^I$, where $\mathcal{H}^I$ is an invariant subspace of $\{M_q\}$, i.e., $\pi^X M_q \pi^I =0$.
Here, $\pi^X$ ($\pi^I$) is the projector of the subspace $\mathcal{H}^X$ ($\mathcal{H}^I$).
Because $A$ is a Hermitian operator, $\pi^X A \pi^X$ can be diagonalized.
According to eigenstates of $\pi^X A \pi^X$, we can further decompose the Hilbert space as $\mathcal{H} = \mathcal{H}_+ \oplus \mathcal{H}_0 \oplus \mathcal{H}_- \oplus \mathcal{H}^I$,
where $\mathcal{H}_\eta$ ($\mathcal{H}_I$) is spanned by $\{\ket{\varphi^{(\eta)}_l}\}$ ($\{\ket{\varphi^I_l}\}$), $\eta=+,0$ and $-$ correspond to positive, zero, and negative eigenvalues of $\pi^X A \pi^X$, respectively.
Then, $A$ can be written as $A = A_+ - A_- + A_I$, where
\begin{equation}
A_{\pm} = \sum_{l} \lambda^{(\pm)}_{l} \ketbra{\varphi^{(\pm)}_l}{\varphi^{(\pm)}_l}
\end{equation}
and
\begin{equation}
A_I = \pi^I A \pi^I + (\pi^I A \pi^{(+)} +\pi^I A \pi^{(0)} +\pi^I A \pi^{(-)} + h.c.).
\end{equation}
Here, $\{\lambda^{(\pm)}_{l}\}$ are all positive, and $\pi^{(\eta)} = \sum_l \ketbra{\varphi^{(\eta)}_l}{\varphi^{(\eta)}_l}$ is the projector of the subspace $\mathcal{H}_\eta$.

\textbf{Lemma 1}. \textit{$\mathcal{H}_\pm$ are two invariant subspaces of $\{M_q\}$, and $A_\pm$ are both MIOs.}

\textit{Proof}. Because $\mathcal{P}$ is a trace preserving CP map, $\tr (\mathcal{P}A_+) = \tr A_+$, where $\tr A_+ = \sum_{l} \lambda^{(+)}_l$ and
\begin{equation}
\tr (\mathcal{P}A_+) = \tr (\pi^{(+)}\mathcal{P}A_+) + \tr (\pi^{(0)}\mathcal{P}A_+)
+ \tr (\pi^{(-)}\mathcal{P}A_+) + \tr (\pi^I\mathcal{P}A_+).
\label{trace}
\end{equation}
Because $A$ is a MIO, $\mathcal{P}A = A$ and $\tr (\pi^{(+)}\mathcal{P}A) = \tr (\pi^{(+)}A)$, where $\tr (\pi^{(+)}A) = \sum_{l} \lambda^{(+)}_l$.
By noticing $\mathcal{H}_I$ is an invariant subspace of $\{M_q\}$ ($\pi^{(+)} M_q \pi^I =0$), we have $\tr (\pi^{(+)}\mathcal{P}A_I) = 0$ and
\begin{equation}
\tr (\pi^{(+)}\mathcal{P}A) = \tr (\pi^{(+)}\mathcal{P}A_+) - \tr (\pi^{(+)}\mathcal{P}A_-).
\label{flow}
\end{equation}
Combining Eqs. (\ref{trace}) and (\ref{flow}), we have
\begin{equation}
\tr (\pi^{(0)}\mathcal{P}A_+) + \tr (\pi^{(-)}\mathcal{P}A_+)
+ \tr (\pi^I\mathcal{P}A_+) = - \tr (\pi^{(+)}\mathcal{P}A_-).
\end{equation}
where each term on the left side is non-negative while the term on the right side is non-positive ($A_+$ and $A_-$ are both positive and $\mathcal{P}$ is a positive map), which implies all terms are zero.
Because
\begin{eqnarray}
\tr (\pi^{(0)}\mathcal{P}A_+) &=& \sum_{q,l,l^\prime} \lambda^{(+)}_l
\abs{\bra{\varphi^{(0)}_{l^{\prime}}} M_q \ket{\varphi^{(+)}_l}}^2 =0 \\
\tr (\pi^{(-)}\mathcal{P}A_+) &=& \sum_{q,l,l^\prime} \lambda^{(+)}_l
\abs{\bra{\varphi^{(-)}_{l^{\prime}}} M_q \ket{\varphi^{(+)}_l}}^2 =0 \\
\tr (\pi^I\mathcal{P}A_+) &=& \sum_{q,l,l^\prime} \lambda^{(+)}_l
\abs{\bra{\varphi^I_{l^{\prime}}} M_q \ket{\varphi^{(+)}_l}}^2 =0,
\end{eqnarray}
we have
\begin{equation}
\bra{\varphi^{(0)}_{l^{\prime}}} M_q \ket{\varphi^{(+)}_l} =
\bra{\varphi^{(-)}_{l^{\prime}}} M_q \ket{\varphi^{(+)}_l} =
\bra{\varphi^I_{l^{\prime}}} M_q \ket{\varphi^{(+)}_l} = 0.
\end{equation}
Similarly,
\begin{equation}
\bra{\varphi^{(+)}_{l^{\prime}}} M_q \ket{\varphi^{(-)}_l} =
\bra{\varphi^{(0)}_{l^{\prime}}} M_q \ket{\varphi^{(-)}_l} =
\bra{\varphi^I_{l^{\prime}}} M_q \ket{\varphi^{(-)}_l} = 0.
\end{equation}
Therefore, $\mathcal{H}_\pm$ are two invariant subspaces of $\{M_q\}$.

Because $\mathcal{H}_I$ and $\mathcal{H}_-$ are invariant subspaces, $\pi^{(+)} M_q \pi^I = \pi^{(+)} M_q \pi^{(-)} =0$.
Thus, $\pi^{(+)} (\mathcal{P}A_I) \pi^{(+)}  = \pi^{(+)} (\mathcal{P}A_-) \pi^{(+)} =0$.
Then, we have $A_+ = \pi^{(+)} A \pi^{(+)} = \pi^{(+)} (\mathcal{P}A) \pi^{(+)} = \pi^{(+)} (\mathcal{P}A_+) \pi^{(+)}$.
Because $\mathcal{H}_+$ is an invariant subspace, $\pi^{(+)} (\mathcal{P}A_+) \pi^{(+)} = \mathcal{P}A_+$.
Therefore, $\mathcal{P}A_+ = A_+$, and $A_+$ is a MIO.
Similarly, $\mathcal{P}A_- = A_-$, and $A_-$ is a MIO.
\hfill $\Box$

Now, we apply Lemma 1 to the case that $\mathcal{H} = \mathcal{H}_X$ and $\mathcal{H}_I$ is empty.
For any Hermitian MIO, positive eigenvalues and negative eigenvalues correspond to two invariant subspaces of $\{M_q\}$, respectively. And, any Hermitian MIO can be written as a linear superposition of two positive Hermitian MIOs.

\subsection{The unique MIO of the map $\mathcal{P}^{(j)}$}

If there exists a Hermitian MIO $\Lambda^{(j)\prime}_R$ which is linearly independent with $\Lambda^{(j)}_R$, one can compose a third nonzero Hermitian MIO $\Lambda^{(j)\prime\prime}_R$ whose trace vanishes, as a linear superposition of $\Lambda^{(j)}_R$ and $\Lambda^{(j)\prime}_R$.
MIO $\Lambda^{(j)\prime\prime}_R$ must have positive and negative eigenvalues.
The map $\mathcal{P}^{(j)}$ is a map in the subsystem $\mathcal{H}^{(j)}_R$.
Now by applying Lemma 1 to the map $\mathcal{P}^{(j)}$, we can find that positive-valued and negative-valued eigenstates of $\Lambda^{(j)\prime\prime}_R$ form two invariant subspaces of $\{ M^{(j)}_q \}$.
However, $\mathcal{H}^{(j)}_{R}$ is irreducible.
Therefore, $\Lambda^{(j)}_R$ is the unique Hermitian MIO up to a scalar factor.

\subsection{The complement subspace}

As shown in the main text, the Hilbert space can be decomposed as $\mathcal{H} = \mathcal{H}^C \oplus \mathcal{H}^{SR}$, where $\mathcal{H}^C$ is the complement subspace and $\mathcal{H}^{SR} = \bigoplus_j (\mathcal{H}^{(j)}_S \otimes \mathcal{H}^{(j)}_R)$ is an invariant subspace of $\{M_q\}$.
Then, we can apply Lemma 1 to the case that $\mathcal{H}^X = \mathcal{H}^C$ and $\mathcal{H}^I = \mathcal{H}^{SR}$.
Without loss of generality, we consider a positive Hermitian MIO.
For a positive Hermitian MIO $A$, eigenvalues of $\pi^C A \pi^C$ must be all zero, otherwise, the complement subspace includes one invariant subspace of $\{M_q\}$ (there is not any negative eigenvalues).
In other words, $\pi^C A \pi^C = 0$.
Here, $\pi^C$ ($\pi^{SR}$) is the projector of the subspace $\mathcal{H}^C$ ($\mathcal{H}^{SR}$).
Because $A$ is positive, all off-diagonal elements between two subspaces $\mathcal{H}^C$ and $\mathcal{H}^{SR}$ are also zero, i.e., $\pi^{SR} A \pi^C = \pi^C A \pi^{SR} = 0$.
Therefore, for any Hermitian MIO $A$, we have $A = \pi^{SR} A \pi^{SR}$ and $\pi^C A = A \pi^C = 0$ (any Hermitian MIO can be written as a linear superposition of two positive MIOs).

\subsection{Off-diagonal elements between two MISs}

In general, we can rewrite the decomposition as $\mathcal{H} = \mathcal{H}^C \oplus \mathcal{H}_1 \oplus \mathcal{H}_2 \oplus \mathcal{H}_3 \oplus \cdots$, where $\mathcal{H}_1$, $\mathcal{H}_2$, $\mathcal{H}_3$, $\ldots$ are MISs.
Because the complement subspace is irrelevant for a Hermitian MIO $A$ ($\pi^C A = A \pi^C = 0$), the Hermitian MIO can be written as $A = \sum_{i,i^\prime} \pi_i A \pi_{i^\prime}$, where $\pi_i$ is the projector of the MIS $\mathcal{H}_i$ and $i=1,2,3,\cdots$.
Because $\{ \mathcal{H}_i \}$ are MISs, $\pi^C M_q \pi_i =0$ and $\pi_i M_q \pi_{i^\prime} =0$ if $i \neq i^\prime$.
Thus, $\pi_i A \pi_{i^\prime} = \pi_i \mathcal{P}(A) \pi_{i^\prime} = \pi_i [\mathcal{P}(\pi_i A \pi_{i^\prime})] \pi_{i^\prime} = \mathcal{P} (\pi_i A \pi_{i^\prime})$, and $\{ \pi_i A \pi_{i^\prime} \}$ are MIOs.
Without loss of generality, we consider two MISs $\mathcal{H}_1$ and $\mathcal{H}_2$.
In the following, we will prove that, if the Hermitian MIO $A_{12} = \pi_1A\pi_2 + \pi_2A\pi_1$ is nonzero, $\mathcal{H}_1$ and $\mathcal{H}_2$ must be isomorphic.
Hence, if $\mathcal{H}_1$ and $\mathcal{H}_2$ are not isomorphic, $\pi_1A\pi_2 = \pi_2A\pi_1 =0$.
Therefore, $\pi^{(j)} A \pi^{(j^\prime)} =0$ if $j \neq j^\prime$.

If the Hermitian MIO $A_{12}$ is nonzero, there must exist two non-empty invariant subspaces $\mathcal{H}_+$ and $\mathcal{H}_-$ corresponding to positive and negative eigenvalues of $A_{12}$, respectively, as a consequence of Lemma 1 ($\mathcal{H}^X = \mathcal{H}_1 \oplus \mathcal{H}_2$ is an invariant subspace of $\{M_q\}$ and $\mathcal{H}^I$ is empty).
Here, we would like to remark that $\tr A_{12} =0$.
For convenience, we denote eigenstates of $A_{12}$ with positive eigenvalues as vectors
$ \{
\left( \begin{array}{c}
\mathbf{u}_l\\
\mathbf{v}_l\\
\end{array} \right)
\} $,
where the vector $\mathbf{u}_l$ ($\mathbf{v}_l$) corresponds to a state in the subspace $\mathcal{H}_1$ ($\mathcal{H}_2$).
In the subspace $\mathcal{H}_1 \oplus \mathcal{H}_2$, measurement operators can be represented as
$
M^{(12)}_q =
\left( \begin{array}{cc}
M^{(1)}_q & 0\\
0 & M^{(2)}_q\\
\end{array} \right)
$,
where $M^{(1)}_q = \pi_1 M_q \pi_1$ and $M^{(2)}_q = \pi_1 M_q \pi_1$ are matrices as the same as measurement operators of corresponding $R$ systems, respectively.

Because $\mathcal{H}_+$ is an invariant subspace of $\{M_q\}$, we have
\begin{equation}
\left( \begin{array}{cc}
M^{(1)}_q & 0\\
0 & M^{(2)}_q\\
\end{array} \right)
\left( \begin{array}{c}
\mathbf{u}_l\\
\mathbf{v}_l\\
\end{array} \right)
= \sum_{l^\prime} \alpha_{l,l^\prime}
\left( \begin{array}{c}
\mathbf{u}_{l^\prime}\\
\mathbf{v}_{l^\prime}\\
\end{array} \right),
\end{equation}
which indicates that $M^{(1)}_q \mathbf{u}_l = \sum_{l^\prime} \alpha_{l,l^\prime} \mathbf{u}_{l^\prime}$ and $M^{(2)}_q \mathbf{v}_l = \sum_{l^\prime} \alpha_{l,l^\prime} \mathbf{v}_{l^\prime}$.
We would like to remark that $\{\mathbf{u}_{l}\}$ and $\{\mathbf{v}_{l}\}$ are decoupled under $M^{(12)}_q$.
Hence, $\{\mathbf{u}_{l}\}$ and $\{\mathbf{v}_{l}\}$ are invariant subspaces of $\{M^{(1)}_q\}$ and $\{M^{(2)}_q\}$, respectively.
The rank of $\{\mathbf{u}_{l}\}$ ($\{\mathbf{v}_{l}\}$) must be the same as the dimension of $\mathcal{H}_1$ ($\mathcal{H}_2$), otherwise, $\mathcal{H}_1$ ($\mathcal{H}_2$) is reducible.
It is similar for the subspace corresponding to negative eigenvalues.
Therefore, the dimensions of $\mathcal{H}_1$, $\mathcal{H}_2$, $\mathcal{H}_+$, and $\mathcal{H}_-$, and the ranks of $\{\mathbf{u}_{l}\}$ and $\{\mathbf{v}_{l}\}$ must be the same.
And $\{\mathbf{u}_{l}\}$ ($\{\mathbf{v}_{l}\}$) is a set of linearly-independent vectors.

Because the ranks of $\{\mathbf{u}_{l}\}$ and $\{\mathbf{v}_{l}\}$ are the same and each of them is a set of linearly-independent vectors, we can define an invertible transformation $T$ satisfying $T\mathbf{u}_l = \mathbf{v}_l$, so that $M^{(1)}_q = T^{-1}M^{(2)}_qT$ and $M^{(2)}_q = TM^{(1)}_qT^{-1}$.
Because $\{ M^{(2)}_q \}$ satisfy the completeness equation, we have
\begin{equation}
\sum_q M^{(2)\dag}_q M^{(2)}_q = \sum_q T^{\dag -1}M^{(1)\dag}_qT^\dag TM^{(1)}_qT^{-1} = \openone_v,
\end{equation}
which means $\sum_q M^{(1)\dag}_q T^\dag T M^{(1)}_q = T^\dag T$, i.e., $T^\dag T$ is a Hermitian invariant operator of the dual map.
Here, $\openone_v$ is the identical operator of the vector space spanned by $\{\mathbf{u}_{l}\}$ (or $\{\mathbf{v}_{l}\}$).
In the next subsection, we will show $T^\dag T$ is proportional to $\openone_v$.
Therefore, $T$ is proportional to a unitary transformation and two subspaces $\mathcal{H}_1$ and $\mathcal{H}_2$ are isomorphic.

\subsection{Dual measurement invariant operators}

A dual map in the MIS $\mathcal{H}_1$ reads $\mathcal{P}^{(1)\dag} \bullet = \sum _q M^{(1)\dag}_q \bullet M^{(1)}_q$.
Because $\{ M^{(1)}_q \}$ satisfy the completeness equation, $\openone_v$ is a dual MIO, i.e., $\sum_q M^{(1)\dag}_q \openone_v M^{(1)}_q = \openone_v$.
If there exists a Hermitian dual MIO that is linearly independent with $\openone_v$, we can show that $\mathcal{H}_1$ is reducible.
Therefore, all dual MIOs of the MIS $\mathcal{H}_1$ are proportional to $\openone_v$.

We suppose $\bar{D}$ is a Hermitian dual MIO that is linearly independent with $\openone_v$.
Then, we always have another nonzero dual MIO $D = \bar{D} - \bar{\lambda}_{min} \openone_v$, where $\bar{\lambda}_{min}$ is the minimal eigenvalue of $\bar{D}$.
The dual MIO $D$ can be written as $D = \sum_{l} \lambda^{(+)}_l \mathbf{w}^{(+)}_l \mathbf{w}^{(+)\dag}_l$, where $\{\lambda^{(+)}_l\}$ are all positive, and $\{ \mathbf{w}^{(+)}_l \}$ ($\{ \mathbf{w}^{(0)}_l \}$) are eigenstates of $D$ with positive (zero) eigenvalues.
We would like to remark that $\{ \mathbf{w}^{(+)}_l \}$ and $\{ \mathbf{w}^{(0)}_l \}$ are both non-empty.
Because $\sum_q M^{(1)\dag}_q D M^{(1)}_q = D$,
\begin{equation}
\sum_l \mathbf{w}^{(0)\dag}_l \sum_q M^{(1)\dag}_q D M^{(1)}_q \mathbf{w}^{(0)}_l
= \sum_{q,l,l^\prime} \lambda^{(+)}_{l^\prime}
\abs{\mathbf{w}^{(+)\dag}_{l^\prime} M^{(1)}_q \mathbf{w}^{(0)}_l}^2 = 0.
\end{equation}
Therefore, $\mathbf{w}^{(+)\dag}_{l^\prime} M^{(1)}_q \mathbf{w}^{(0)}_l = 0$ and $\{ \mathbf{w}^{(0)}_l \}$ is an invariant subspace of $\{M^{(1)}_q\}$.

\section{The effective Hamiltonian}

Because $\{ \mathcal{H}^{(j)}_S \otimes \mathcal{H}^{(j)}_R \}$ are invariant subspaces of $\{M_q\}$,
\begin{equation}
M_q = M_q \pi^C + \sum_j \pi^{(j)} M_q \pi^{(j)}.
\label{MOperator}
\end{equation}
Then,
\begin{equation}
M_q^\dag M_q = \pi^C M_q^\dag M_q \pi^C + \sum_j \pi^C M_q^\dag \pi^{(j)} M_q \pi^{(j)}
+\sum_j \pi^{(j)} M_q^\dag \pi^{(j)} M_q \pi^C + \sum_j \pi^{(j)} M_q^\dag \pi^{(j)} M_q \pi^{(j)}.
\end{equation}
Due to the completeness equation, we have
\begin{eqnarray}
\sum_q \pi^C M_q^\dag M_q \pi^C &=& \pi^C, \\
\sum_q \pi^{(j)} M_q^\dag \pi^{(j)} M_q \pi^{(j)} &=& \pi^{(j)}, \label{jj}
\end{eqnarray}
and
\begin{equation}
\sum_q \pi^C M_q^\dag \pi^{(j)} M_q \pi^{(j)} = \sum_q \pi^{(j)} M_q^\dag \pi^{(j)} M_q \pi^C =0.
\label{cjjc}
\end{equation}

\textbf{Lemma 2}. \textit{For any operator $A$, if $\pi^C A \pi^C = 0$ and $\tr_R (\pi^{(j)} A \pi^{(j)}) = \tilde{A}^{(j)}$, $\pi^C \mathcal{P}(A) \pi^C = 0$ and $\tr_R [\pi^{(j)} \mathcal{P}(A) \pi^{(j)}] = \tilde{A}^{(j)}$.}

\textit{Proof}. Using Eq. (\ref{MOperator}), we have
\begin{equation}
\pi^C \mathcal{P}(A) \pi^C = \sum_q \pi^C M_q A M_q^\dag \pi^C
= \sum_q \pi^C M_q \pi^C A \pi^C M_q^\dag \pi^C =0,
\end{equation}
and
\begin{eqnarray}
\pi^{(j)} \mathcal{P}(A) \pi^{(j)} &=& \sum_q \pi^{(j)} M_q A M_q^\dag \pi^{(j)} \notag \\
&=& \sum_q \pi^{(j)} M_q \pi^{(j)} A \pi^{(j)} M_q^\dag \pi^{(j)}
+ \pi^{(j)} M_q \pi^C A \pi^{(j)} M_q^\dag \pi^{(j)} + \pi^{(j)} M_q \pi^{(j)} A \pi^C M_q^\dag \pi^{(j)}.
\end{eqnarray}
By noticing $\pi^{(j)} M_q \pi^{(j)} = \openone^{(j)}_S \otimes M^{(j)}_q$ and using Eqs. (\ref{jj}) and (\ref{cjjc}), we have
\begin{equation}
\tr_R (\sum_q \pi^{(j)} M_q \pi^{(j)} A \pi^{(j)} M_q^\dag \pi^{(j)})
= \tr_R (\sum_q \pi^{(j)} M_q^\dag \pi^{(j)} M_q \pi^{(j)} A \pi^{(j)})
= \tr_R (\pi^{(j)} A \pi^{(j)}) =\tilde{A}^{(j)}
\end{equation}
and
\begin{eqnarray}
\tr_R (\sum_q \pi^{(j)} M_q \pi^C A \pi^{(j)} M_q^\dag \pi^{(j)})
&=& \tr_R (\sum_q \pi^{(j)} M_q^\dag \pi^{(j)} M_q \pi^C A \pi^{(j)}) =0, \\
\tr_R (\sum_q \pi^{(j)} M_q \pi^{(j)} A \pi^C M_q^\dag \pi^{(j)})
&=& \tr_R (\sum_q \pi^{(j)} A \pi^C M_q^\dag \pi^{(j)} M_q \pi^{(j)}) =0. \\
\end{eqnarray}
Here, we have used that for any operator $X$ and operator $Y_R$ in the subsystem $R$, $\tr_R [X (\openone_S \otimes Y_R)] = \tr_R [(\openone_S \otimes Y_R) X]$.
Therefore, $\tr_R [\pi^{(j)} \mathcal{P}(A) \pi^{(j)}] = \tilde{A}^{(j)}$.
\hfill $\Box$

\textbf{Lemma 3}. \textit{For an operator $A$, if $\pi^C A \pi^C = 0$ and $\tr_R (\pi^{(j)} A \pi^{(j)}) = \tilde{A}^{(j)}$, $\tr_R [\pi^{(j)} \mathcal{S}_\infty (A) \pi^{(j)}] = \tilde{A}^{(j)}$ and $\pi^{(j)} \mathcal{S}_\infty (A) \pi^{(j)} = \tilde{A}^{(j)} \otimes \Lambda^{(j)}_R$.}

\textit{Proof}. Using Lemma 2, we have, $\pi^C \mathcal{P}^m(A) \pi^C = 0$ and $\tr_R [\pi^{(j)} \mathcal{P}^m(A) \pi^{(j)}] = \tilde{A}^{(j)}$ for any $m$.
Hence, $\tr_R [\pi^{(j)} \mathcal{S}_\infty (A) \pi^{(j)}] = \tilde{A}^{(j)}$.
Because $\mathcal{S}_\infty (A)$ is a MIO and $\tr \Lambda^{(j)}_R = 1$, $\pi^{(j)} \mathcal{S}_\infty (A) \pi^{(j)} = \tilde{A}^{(j)} \otimes \Lambda^{(j)}_R$.
\hfill $\Box$

\textbf{Effective Hamiltonian}. If the state $\rho$ is a MIO, $\rho \pi^C =0$.
Hence, $\pi^C H\rho \pi^C =0$.
Using Lemma 3, we have $\pi^{(j)} \mathcal{S}_\infty (H\rho) \pi^{(j)} = \widetilde{H\rho}^{(j)} \otimes \Lambda^{(j)}_R$, where $\widetilde{H\rho}^{(j)} = \tr_R (\pi^{(j)} H\rho \pi^{(j)})$.

We suppose the MIO $\rho = \bigoplus_j (\rho^{(j)}_S \otimes \Lambda^{(j)}_R)$, and $\pi^{(j)} H \pi^{(j)} = \sum_{s,s^\prime} \ketbra{\Phi^{(j)}_s}{\Phi^{(j)}_{s^\prime}} \otimes H^{(j)}_{s,s^\prime}$.
Then, $\rho \pi^{(j)} = \pi^{(j)} \rho \pi^{(j)} = (\rho^{(j)}_S \otimes \Lambda^{(j)}_R)$, and
\begin{equation}
\tr_R (\pi^{(j)} H\rho \pi^{(j)}) = \tr_R [\pi^{(j)} H \pi^{(j)} (\rho^{(j)}_S \otimes \Lambda^{(j)}_R)]
= \sum_{s,s^\prime} \tr_R(H^{(j)}_{s,s^\prime}\Lambda^{(j)}_R)
\ketbra{\Phi^{(j)}_s}{\Phi^{(j)}_{s^\prime}} \rho^{(j)}_S.
\end{equation}
Therefore, $\tr_R[\pi^{(j)} \mathcal{S}_\infty (H\rho) \pi^{(j)}] = \tilde{H}^{(j)}_S \rho^{(j)}_S$,
where
\begin{equation}
\tilde{H}^{(j)}_S = \sum_{s,s^\prime} \tr_R(H^{(j)}_{s,s^\prime}\Lambda^{(j)}_R)
\ketbra{\Phi^{(j)}_s}{\Phi^{(j)}_{s^\prime}}
= \tr_R [ \pi^{(j)} H \pi^{(j)} (\openone^{(j)}_S \otimes \Lambda^{(j)}_R) ].
\end{equation}

Because $\mathcal{S}_\infty (H\rho)$ is a MIO, $\mathcal{S}_{\infty}H\rho = \tilde{H}\rho$, where $\tilde{H} = \bigoplus_j (\tilde{H}^{(j)}_S \otimes \openone^{(j)}_R)$.
Similarly, $\mathcal{S}_{\infty}\rho H = \rho \tilde{H}$.

\section{The proof of the Zeno effect with non-selective measurements}

As we will show in the following, $\Delta$ includes three parts for each time interval of $\tau/N_{2}$, and
\begin{equation}
\Delta = \sum _{n=1}^{N_2} [ \Delta_{\text{I}}(n) + \Delta_{\text{II}}(n) + \Delta_{\text{III}}(n) ].
\end{equation}
By using the notation $\mathcal{V}_{N_1} = [ \mathcal{P} \mathcal{U}(\tau/N) ]^{N_1}$, we have
\begin{eqnarray}
&& \rho(\tau) = \mathcal{V}_{N_1}^{N_2} \rho(0) \notag \\
&=& \mathcal{V}_{N_1}^{N_2-1} e^{\tilde{\mathcal{L}} \tau/N_2} \rho(0)
+ \Delta_{\text{I}}(1) + \Delta_{\text{II}}(1) + \Delta_{\text{III}}(1) \notag \\
&=& \mathcal{V}_{N_1}^{N_2-2} e^{\tilde{\mathcal{L}} 2\tau/N_2} \rho(0)
+ \Delta_{\text{I}}(1) + \Delta_{\text{II}}(1) + \Delta_{\text{III}}(1)
+ \Delta_{\text{I}}(2) + \Delta_{\text{II}}(2) + \Delta_{\text{III}}(2) \notag \\
&\ldots & \notag \\
&=& e^{\tilde{\mathcal{L}} \tau} \rho(0)
+ \sum _{n=1}^{N_2} [ \Delta_{\text{I}}(n) + \Delta_{\text{II}}(n) + \Delta_{\text{III}}(n) ].
\end{eqnarray}
For each time interval of $\tau/N_{2}$,
\begin{eqnarray}
&&\mathcal{V}_{N_1}^{N_2-n+1} e^{\tilde{\mathcal{L}} (n-1)\tau/N_2} \rho(0) \notag \\
&=& \mathcal{V}_{N_1}^{N_2-n} [1 + (\tau/N_2)\mathcal{S}_{N_1} \mathcal{L}] e^{\tilde{\mathcal{L}} (n-1)\tau/N_2} \rho(0)
+ \Delta_{\text{I}}(n) \notag \\
&=& \mathcal{V}_{N_1}^{N_2-n} [1 + (\tau/N_2)\tilde{\mathcal{L}}] e^{\tilde{\mathcal{L}} (n-1)\tau/N_2} \rho(0)
+ \Delta_{\text{I}}(n) + \Delta_{\text{II}}(n) \notag \\
&=& \mathcal{V}_{N_1}^{N_2-n} e^{\tilde{\mathcal{L}} n\tau/N_2} \rho(0)
+ \Delta_{\text{I}}(n) + \Delta_{\text{II}}(n) + \Delta_{\text{III}}(n).
\end{eqnarray}
Here,
\begin{equation}
\Delta_{\text{I}}(n) = \mathcal{V}_{N_1}^{N_2-n}
\{ \mathcal{V}_{N_1} - [1 + (\tau/N_2)\mathcal{S}_{N_1} \mathcal{L}] \}
e^{\tilde{\mathcal{L}} (n-1)\tau/N_2} \rho(0),
\end{equation}
\begin{equation}
\Delta_{\text{II}}(n) = \mathcal{V}_{N_1}^{N_2-n}
\{ [1 + (\tau/N_2)\mathcal{S}_{N_1} \mathcal{L}] - [1 + (\tau/N_2)\tilde{\mathcal{L}}] \}
e^{\tilde{\mathcal{L}} (n-1)\tau/N_2} \rho(0)
\end{equation}
and
\begin{equation}
\Delta_{\text{III}}(n) = \mathcal{V}_{N_1}^{N_2-n}
\{ [1 + (\tau/N_2)\tilde{\mathcal{L}}] - e^{\tilde{\mathcal{L}} \tau/N_2} \}
e^{\tilde{\mathcal{L}} (n-1)\tau/N_2} \rho(0).
\end{equation}

\subsection{The norm of $\Delta_{\text{I}}(n)$}

As shown in the main text, $\rho_n = e^{\tilde{\mathcal{L}} (n-1)\tau/N_2} \rho(0)$ is a MIO.
Thus,
\begin{equation}
\Delta_{\text{I}}(n) = \mathcal{V}_{N_1}^{N_2-n}
\left(
\mathcal{V}_{N_1}
-
[\mathcal{P}^{N_1} + (\tau/N)\sum _{m=1}^{N_1} \mathcal{P}^{m} \mathcal{L} \mathcal{P}^{(N_1-m)} ]
\right)
\rho_n,
\end{equation}
Because unitary operations ($\mathcal{U}$) and trace preserving CP maps ($\mathcal{P}$) do not increase the trace norm of a Hermitian operator (see the last paragraph of this subsection for explanation), $\mathcal{V}_{N_1}$ do not increase the trace norm of a Hermitian operator, and we have
\begin{equation}
\norm{\Delta_{\text{I}}(n)}_1
\leq \left\|
\left(
[\mathcal{P} \mathcal{U}(\tau/N)]^{N_1}
-
[\mathcal{P}^{N_1} + (\tau/N)\sum _{m=1}^{N_1} \mathcal{P}^{m} \mathcal{L} \mathcal{P}^{(N_1-m)} ]
\right) \rho_n
\right\|_1 .
\end{equation}
After expanding evolution operators, we have
\begin{equation}
\norm{\Delta_{\text{I}}(n)}_1
\leq \left\|
\left(
[\mathcal{P} \sum_{l=0}^{\infty} \frac{(\tau/N)^l}{l!}\mathcal{L}^l]^{N_1}
- [\mathcal{P}^{N_1} + (\tau/N)\sum _{m=1}^{N_1} \mathcal{P}^{m} \mathcal{L} \mathcal{P}^{(N_1-m)} ]
\right) \rho_n
\right\|_1 ,
\label{expansion}
\end{equation}
where terms of the second part are all included in the expansion of the first part (corresponding to the term without $\mathcal{L}$ and terms with only one $\mathcal{L}$ of the first part).
After further expanding,
\begin{equation}
\norm{\Delta_{\text{I}}(n)}_1
\leq \sum_{\{n_i\}} \sum_{\{m_i\}} \alpha_{\{n_i\}\{m_i\}}
\norm{
\mathcal{P}^{m_{N_1}} \mathcal{L}^{n_{N_1}} \cdots \mathcal{P}^{m_2} \mathcal{L}^{n_2} \mathcal{P}^{m_1} \mathcal{L}^{n_1} \mathcal{P}^{m_0}
\rho_n }_1,
\end{equation}
where $\{n_i\}$ and $\{m_i\}$ are some strings of non-negative integers ($\sum_i n_i \geq 2$ and $\sum_i m_i = N_1$) and $\{ \alpha_{\{n_i\}\{m_i\}} \}$ are all positive real coefficients.
Again, because trace preserving CP maps do not increase the trace norm of a Hermitian operator, we have
\begin{equation}
\norm{\Delta_{\text{I}}(n)}_1 \leq
\sum_{\{n_i\}} \sum_{\{m_i\}} \alpha_{\{n_i\}\{m_i\}} (2J)^{\sum_i n_i} \norm{\rho_n}_1,
\end{equation}
where the right side can be obtained by replacing $\mathcal{P}$ with $1$, $\mathcal{L}$ with $2J$, and $\rho_n$ with $\norm{\rho_n}_1$ in the right side of Eq. (\ref{expansion}), i.e.,
\begin{equation}
[ e^{2J\tau/N_2} - (1 + 2J\tau/N_2) ] \norm{\rho_n}_1
= \sum_{\{n_i\}} \sum_{\{m_i\}} \alpha_{\{n_i\}\{m_i\}} (2J)^{\sum_i n_i} \norm{\rho_n}_1.
\end{equation}
Because $\norm{\rho_n}_1 = 1$, we have
\begin{equation}
\norm{\Delta_{\text{I}}(n)}_1 \leq [ e^{2J\tau/N_2} - (1 + 2J\tau/N_2) ].
\end{equation}

\textbf{Trace norm and the trace preserving CP map}. For a Hermitian operator, the trace norm is the sum of the absolute values of eigenvalues.
A Hermitian operator $A$ can be decomposed as $A = A_+ - A_-$, where $A_+$ and $A_-$ are two positive Hermitian operators corresponding to positive eigenvalues and negative eigenvalues of $A$, respectively.
Then,  $\norm{A_\pm}_1=\tr A_\pm$ and $\norm{A}_1 = \tr(A_+ + A_-)$.
Because $\mathcal{P}A_\pm$ are also positive Hermitian operators, $\norm{\mathcal{P}A}_1 \leq \norm{\mathcal{P}A_+}_1 + \norm{\mathcal{P}A_-}_1 = \tr [\mathcal{P}(A_+ + A_-)]  = \norm{A}_1$.

\subsection{The norm of $\Delta_{\text{II}}(n)$}

It is straightforward that
\begin{equation}
\norm{\Delta_{\text{II}}(n)}_1 \leq
(\tau/N_2) \norm{(\mathcal{S}_{N_1} \mathcal{L} -\tilde{\mathcal{L}}) \rho_n}_1.
\end{equation}

\subsection{The norm of $\Delta_{\text{III}}(n)$}

Similar to $\Delta_{\text{I}}(n)$, after expanding, one can find that
\begin{equation}
\norm{\Delta_{\text{III}}(n)}_1 \leq
[ e^{2\tilde{J}\tau/N_2} - (1 + 2\tilde{J}\tau/N_2) ].
\end{equation}

\subsection{The norm of $\Delta$}

In summary,
\begin{eqnarray}
\norm{\Delta}_1 &\leq &
\sum _{n=1}^{N_2} \norm{\Delta_{\text{I}}(n)}_1+\norm{\Delta_{\text{II}}(n)}_1+\norm{\Delta_{\text{III}}(n)}_1
\notag \\ &\leq &
N_2 \{ [e^{2J\tau/N_2} - (1 + 2J\tau/N_2)] + [ e^{2\tilde{J}\tau/N_2} - (1 + 2\tilde{J}\tau/N_2) ] \}
+(\tau/N_2) \sum_{n=1}^{N_2} \norm{(\mathcal{S}_{N_1} \mathcal{L} -\tilde{\mathcal{L}}) \rho_n}_1.
\end{eqnarray}

\section{The proof of the Zeno effect with selective measurements}

Firstly, we consider an an initial state that is a product state of two subsystems, e.g., $\rho (0) = \rho^{(j)}_S(0) \otimes \Lambda^{(j)}_R$.
Without loss of generality, we suppose $\rho^{(j)}_S(0) = \sum_s w_s \ketbra{\Phi^{(j)}_s}{\Phi^{(j)}_s}$.
By introducing a virtual system $\mathcal{H}^{(j)}_{\bar{S}}$ spanned by $\{\ket{\bar{\Phi}^{(j)}_s}\}$, the state $\rho^{(j)}_S(0)$ can be represented as the reduced state of a pure state $\ket{\Psi(0)}=\sum_s \sqrt{w_s} \ket{\bar{\Phi}^{(j)}_s} \otimes \ket{\Phi^{(j)}_s}$ in the Hilbert $\mathcal{H}^{(j)}_{\bar{S}} \otimes \mathcal{H}^{(j)}_S$, i.e., $\rho^{(j)}_S(0)=\tr_{\bar{S}}\ketbra{\Psi(0)}{\Psi(0)}$.
Then, the initial state in the extended Hilbert space $\mathcal{H}^{(j)}_{\bar{S}} \otimes \mathcal{H}^{(j)}_S \otimes \mathcal{H}^{(j)}_R$ is $\rho_{\text{ext.}} (0) = \ketbra{\Psi(0)}{\Psi(0)} \otimes \Lambda^{(j)}_R$.

For non-selective measurements, the final state in the extended Hilbert space is $\rho_{\text{ext.}} (\tau) = \ketbra{\Psi(\tau)}{\Psi(\tau)} \otimes \Lambda^{(j)}_R$, where $\ket{\Psi(\tau)} = e^{-i\openone_{\bar{S}}\otimes \tilde{H}^{(j)}_S \tau} \ket{\Psi(0)}$ and $\openone_{\bar{S}}$ is the identity operator of the virtual subsystem.
And the state $\ket{\Psi(\tau)}$ satisfies $\tr_{\bar{S}}\ketbra{\Psi(\tau)}{\Psi(\tau)} = \rho^{(j)}_S(\tau) = e^{-i\tilde{H}^{(j)}_S \tau} \rho^{(j)}_S(0) e^{i\tilde{H}^{(j)}_S \tau}$.

For selective measurements, we suppose the final state in the extended Hilbert space is $\rho_{\text{ext.}} (\tau; \{ q \})$.
The final states for non-selective measurements and selective measurements have the relation $\rho_{\text{ext.}} (\tau) = \sum_{\{ q \}} \rho_{\text{ext.}} (\tau; \{ q \})$.
Here, states $\rho_{\text{ext.}} (\tau; \{ q \})$ are not normalized.
Hence, $\ketbra{\Psi(\tau)}{\Psi(\tau)} = \sum_{\{ q \}} \tr_R \rho_{\text{ext.}} (\tau; \{ q \})$.
Because $\ketbra{\Psi(\tau)}{\Psi(\tau)}$ is a pure state, $\tr_R \rho_{\text{ext.}} (\tau; \{ q \}) \propto \ketbra{\Psi(\tau)}{\Psi(\tau)}$ for any outcomes, i.e., $\rho_{\text{ext.}} (\tau; \{ q \}) = \ketbra{\Psi(\tau)}{\Psi(\tau)} \otimes \rho^{(j)}_R(\{ q \})$.
Using $\rho (\tau; \{ q \}) = \tr_{\bar{S}}\rho_{\text{ext.}} (\tau; \{ q \})$, one find that for the product-state initial state, $\rho (\tau; \{ q \}) = \rho^{(j)}_S(\tau) \otimes \rho^{(j)}_R(\{ q \})$.

In general, a MIO initial state is a linear superposition of product-state initial states, i.e., $\rho(0) = \bigoplus_j (\rho^{(j)}_S(0) \otimes \Lambda^{(j)}_R)$.
Then, the final state for selective measurements is also a superposition of $\rho^{(j)}_S(\tau) \otimes \rho^{(j)}_R(\{ q \})$, i.e., $\rho (\tau; \{ q \}) = \bigoplus_j [\rho^{(j)}_S(\tau) \otimes \rho^{(j)}_R(\{ q \})]$.

\end{document}